\documentclass[conference]{IEEEtran}
\IEEEoverridecommandlockouts

\usepackage{cite}
\usepackage{amsmath,amssymb,amsfonts}
\usepackage{graphicx}
\usepackage{textcomp}
\usepackage{xcolor}

\usepackage{booktabs,threeparttable}
\usepackage{subcaption} 
\usepackage{algorithm}

\usepackage[noend]{algpseudocode}
\usepackage{comment} 
\usepackage{url}
\providecommand{\Description}[2][]{}

\def\BibTeX{{\rm B\kern-.05em{\sc i\kern-.025em b}\kern-.08em
    T\kern-.1667em\lower.7ex\hbox{E}\kern-.125emX}}
\begin{document}

\title{LLM-based Zero-shot Triple Extraction for Automated Ontology Generation from Software Engineering Standards}

\author{\IEEEauthorblockN{1\textsuperscript{st} Songhui Yue}
\IEEEauthorblockA{\textit{Computer Science Department} \\
\textit{Charleston Southern University}\\
North Charleston, USA \\
syue@csuniv.edu}
}
\maketitle

\begin{abstract}
Ontologies have supported knowledge representation and white-box reasoning for decades; thus, the automated ontology generation (AOG) plays a crucial role in scaling their use. Software engineering standards (SES) consist of long, unstructured text (with high noise) and paragraphs with domain-specific terms. In this setting, relation triple extraction (RTE), together with term extraction, constitutes the first stage toward AOG. This work proposes an open-source large language model (LLM)-assisted approach to RTE for SES. Instead of solely relying on prompt-engineering-based methods, this study promotes the use of LLMs as an aid in constructing ontologies and explores an effective AOG workflow that includes document segmentation, candidate term mining, LLM-based relation inference, term normalization, and cross-section alignment. Expert-annotated reference sets at three granularities are constructed and used to evaluate the ontology generated from the study. The results show that it is comparable and potentially superior to the OpenIE method of triple extraction. 
\end{abstract}


\section{Introduction}

\begin{figure*}[h]
	\centering
	\includegraphics[width=0.9\linewidth]{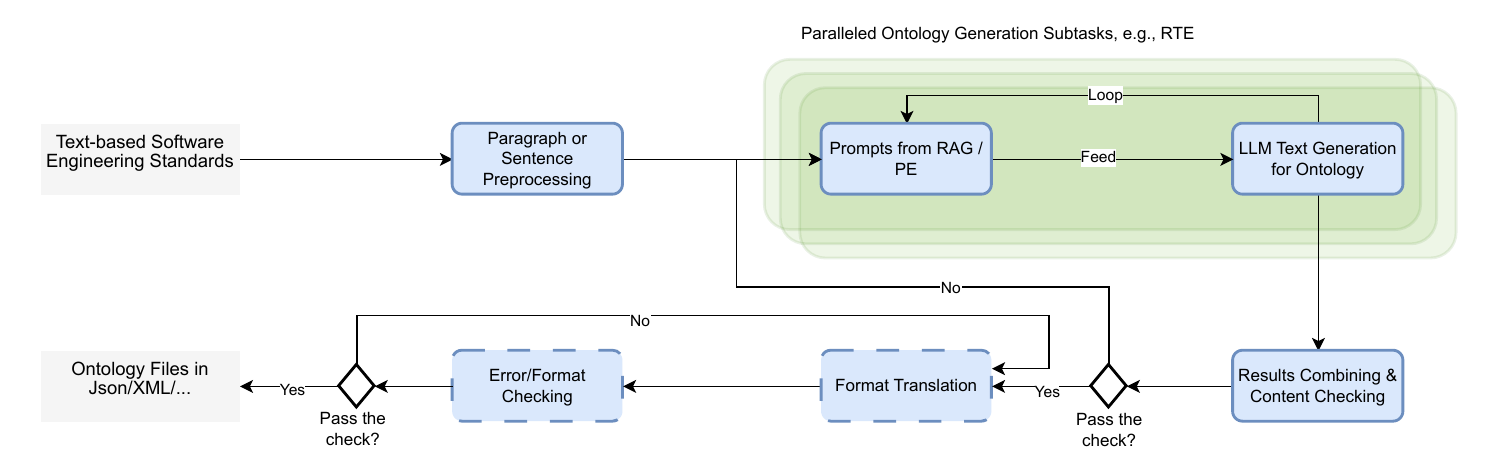}
	\caption{A Workflow Overview for LLM-Assisted Ontology Generation}
	\label{workflow}
\end{figure*}

Automated ontology learning has been discussed since 2007 \cite{bedini2007automatic}, but only in recent years has research on automated ontology generation (AOG) emerged \cite{al2020automatic, navarro2020automated, elnagar2022automatic, leshcheva2022method}. Traditionally, ontologies are created manually \cite{omelayenko2001learning}, semi-automatically from text \cite{maedche2000semi}, or from structured/tabular data such as relational databases \cite{lakzaei2021ontology}. Recent advancements in automated ontology generation leverage deep learning methods \cite{navarro2020automated, al2020automatic}. Originally text-centric in natural language processing (NLP), LLMs now function as general-purpose models that handle multimodal inputs \cite{karanikolas2025strengths, zhang2024mm} and thus are selected to serve as the main assistance for this AOG study. 

It is evident that LLMs have been able to help ontology generation from text \cite{babaei2023llms4ol, kommineni2024human, neuhaus2023ontologies}. When prompted with a passage, an LLM can be instructed to output ontology artifacts in a designated schema (e.g., triples in JSON or OWL/Turtle). However, naive prompting for AOG faces three challenges: 1) Reproducibility and prompt sensitivity: sometimes, when the prompt changes or there is no standard prompt format, the generated content can vary. Even using the same prompt, the generated result (e.g., terms and relations) will vary. 2) Scaling to long or multimodal sources: the generation involving processing a large corpus of text, or multimodal information(such as text and images) requires cross-section term consolidation. A large corpus of text will include many chapters and paragraphs; thus, ontology alignment will need to be considered so that the terms will be able to match the terms in another section. 3) Engineering workflow. When the problem size is large and requires ongoing maintenance and evolution, a mature workflow is needed for the complex task of AOG.

Software engineering standards (SES) may have been implicitly encoded in LLM in the form of model parameters during pretraining; however, to the best of our knowledge, the effort of explicitly representing and extracting that knowledge in a structured way, such as in an ontology, which supports knowledge reasoning \cite{yue2021applying, yue2023csm} and human-centered/human-in-the-loop knowledge representation \cite{opasjumruskit2022ontohuman, tsaneva2024enhancing}, has not been thoroughly investigated. Due to the complexity involved in the tasks of AOG for SES, the problem of this study is deliberately scoped as follows: 1) A single SES and its official short version are used to keep the text focused and easier to build evaluation benchmarks. 2) It is concentrated on relation triple extraction (RTE) as the first stage toward an LLM-based, fully automated ontology-generation pipeline whose goals are usability, faithfulness to the source text, and maximal coverage. 3) It employs open-source LLMs (e.g., Mistral-7B) and encoder-only models (e.g., BERT) due to cost control and potential privacy concerns \cite{kibriya2024privacy} in the usage of such methods.

Compared with prevalent TBox-prioritized methodologies (first identifying types and taxonomy) for ontology creation \cite{babaei2023llms4ol, kommineni2024human, neuhaus2023ontologies}, this study proposes an assertion-led, ABox (instance-level assertion)–TBox co-extraction AOG approach for processing textual SES, designs an engineering workflow and best practices for utilizing open-source LLM, and explores metrics for article data sets that are noisy and have open relations. The subsequent sections are organized as follows: Section \ref{sec:rw} reviews related work. Section \ref{sec:method} details the methodology. Section \ref{sec:experiments} presents experiments and preliminary results. Section \ref{sec:conclusion} discusses limitations and future work and concludes.

\section{Related Work}
\label{sec:rw}
Giglou et al. \cite{babaei2023llms4ol} present LLMs4OL, a conceptual framework that spans tasks across TBox and ABox for ontology learning with LLMs. In their experimental emphasis—and in Lo et al. \cite{lo2024end}—the approach is TBox-first, starting with type recognition and taxonomy induction that is limited primarily to is-a hierarchies. In contrast, this study emphasize extracting entity–relation assertions (triples) from SES text, and future work will focus on deriving concepts and their hierarchies from these assertions. Prior evaluations have been conducted on general domain resources (e.g., WordNet), geographic resources (e.g., GeoNames), and biomedical resources (e.g., NCI), and assess a mix of commercial and open-source LLMs. Our work targets SES—an open-relation setting with higher noise and scarce gold ontologies—and builds an engineering workflow exclusively with open-source LLMs. Related work by Lippolis et al. \cite{lippolis2025ontology} focuses on prompt engineering (PE), proposing two improved prompting techniques for ontology generation.

This work \cite{kommineni2024human} works on a semi-automated workflow for knowledge graph construction, assisted by LLMs (such as ChatGPT 3.5). Human-in-the-loop is involved in different stages of their LLM-supported pipeline. Their approach relies on collecting competency questions to create an ontology and to fill the data in the ontology. They apply their methodology to creating an ontology and KGs about deep learning (DL) methodologies extracted from scholarly publications in the biodiversity domain.

Gong et al. \cite{gong2021prompt} propose ZS-SKA, an approach for prompt-based zero-shot RTE, and apply semantic knowledge augmentation to recognize unseen relations. Their data augmentation is through translating a sentence from its original seen relation to a new unseen relation using an analogy. In the word level, they use top-10 similar words as candidates for new words. Like our work, theirs operates in a zero-shot setting; yet, their datasets (e.g., FewRel) for evaluation are sentence-level with predefined relation sets  \cite{han2018fewrel}. In contrast, our study targets SES with open-set relations, article-level with multiple paragraphs and higher noise, and focuses on extracting entity–relation triples. 

\section{Methodology}
\label{sec:method}

\subsection{Workflow Overview}
Compared with TBox-prioritized methods (first identifying types and taxonomy), this study propose an assertion-led, ABox–TBox coextraction pipeline: from SES text, relation triplet extraction (RTE) yields instance-level assertions and schema-level candidates (e.g., classes, is-a, domain/range), which could be later validated and consolidated into an OWL ontology. The workflow shown in Figure \ref{workflow} consists of two main phases, the upper components of which focus on the subtasks of ontology generation using LLMs. The lower components are for postprocessing, including result combination and various checks, which are planned for the next step.

The subtasks shown in the green rectangle are executed in parallel, allowing some processes to run simultaneously. Those generations can be triples or relations, or even fixing the quality issues according to the content checking results.

\subsection{Standard Selection for this Study}
Among many available software engineering standards, the "Software Engineering Code of Ethics and Professional Practice" (SECEPP) is selected for initial exploration. The main reasons for choosing this document include:
1) It is a publicly available software engineering standard and "is meant to be a useful code, a document that can inform practice and education" \cite{acm1999code}.
2) It is concise and not too long, but long enough to be used for starting the investigation.
3) It has a short version (278 words) and a full version (2653 words), where the short version can be used to do concept verification, and the full version is suitable for testing and further investigation. The organization of the chapters and lists imports common noise and challenges of SES processing, which facilitate the migration of this method to work on other types of documents.
4) For automated software engineering and evolution, it is essential that those agents comply with the ethics provided in this article. Thus, it is significant to have it in ontology form/ structured form.

\subsection{Building an Ontology Scaffold}
One challenge of prompt engineering for open-source LLMs is to generate a stable format of the output. The performance of the LLMs could be totally different for different required output formats (e.g., a simple array or JSON triplet format). It may even decide whether the whole route will be successful or not. Thus, instead of relying on LLMs to generate everything, this study primarily relies on the capability of LLMs to find the relations of nodes (terms). As a result, a concept–relation graph (triples) is automatically extracted from the code text. This graph serves as an ontology scaffold for subsequent enrichment.

Algorithm 1 in pseudo code describes the process of building an ontology scaffold \(G=(V, E)\) from SECEPP by processing text sentence-by-sentence. For each sentence \(S\), spaCy \cite{honnibal2020spacy} extracts noun phrases and verbs to form candidate entities \(P\) and relation vocabulary \(V_b\); both are cleaned and deduplicated. If no verb candidates exist, the sentence is saved as an orphan \(O\). A constrained prompt is composed with \((S, P, V_b)\); a low-temperature LLM returns JSON triples \(T\) under a dynamic token budget (calculated according to the number of terms of the sentence). The method strictly parses the JSON (retry up to \(k=3\)), normalizes each term (e.g., engineers to engineer, running to run), adds nodes to \(V\), and edges to \(E\). The lightweight term normalization is via spaCy lemmatization (e.g., engineers → engineer), preserving acronyms/proper nouns and rejecting overly aggressive changes using a similarity threshold. After processing all sentences, remaining orphans are inserted into \(V\). The result is a low-noise scaffold ready for later typing/merging. Figure \ref{fig:ontology-panels} panel (d) is an exemplified scaffold generated by the automated workflow of this study.

\begin{algorithm}[t]
	\caption{Building an Ontology Scaffold through RTE}
	\label{alg:llm-onto}
	\small
	\begin{algorithmic}[1]   
		\Require SES Article $A$; spaCy model $\mathit{nlp}$; LLM $\mathit{llm}$
		\Ensure $G=(\mathcal{V},\mathcal{E})$
		\State $\mathcal{V}\gets\emptyset$; $\mathcal{E}\gets\emptyset$; $\mathcal{O}\gets\emptyset$
		\For{paragraph $p$ in $A$}
		\State $S\gets \textsc{SplitIntoSentences}(p)$
		\For{sentence $s$ in $S$}
		\State $P\gets \textsc{CleanDedup}(\textsc{ExtractNounPhrases}(s))$
		\State $V_b\gets \textsc{CleanDedup}(\textsc{ExtractVerbs}(s))$
		\If{$P=\emptyset$ \textbf{and} $V_b=\emptyset$} \State \textbf{continue} \EndIf
		\State $M\gets \textsc{DynamicMaxTokens}(P,V_b,256,1024)$
		\State $prompt\gets \textsc{ComposeConstrainedPrompt}(s,P,V_b)$
		\State $T\gets \textsc{Retry}_{k=3}\big(\textsc{ParseValidJSON}(llm(prompt,\allowbreak T{=}0.2,\ \texttt{max\_new\_tokens}{=}M))\big)$
		\If{$T\neq\emptyset$}
		\For{$\langle subj,pred,obj\rangle \in T$}
		\State $subj\gets \textsc{NormalizeTerm}(subj)$
		\State $obj\gets \textsc{LemmatizeTerm}(obj)$
		\State $\mathcal{E}\gets \mathcal{E}\cup\{\langle subj,pred,obj\rangle\}$
		\State $\mathcal{V}\gets \mathcal{V}\cup\{subj,obj\}$
		\EndFor
		\EndIf
		\EndFor
		\EndFor
		\State \textbf{return} $G$
	\end{algorithmic}
\end{algorithm}

\subsection{Data and Expert-annotated Reference Sets}
\begin{figure*}[t]
	\centering
	\begin{subfigure}[b]{0.32\textwidth}
		\centering
		\includegraphics[width=\linewidth]{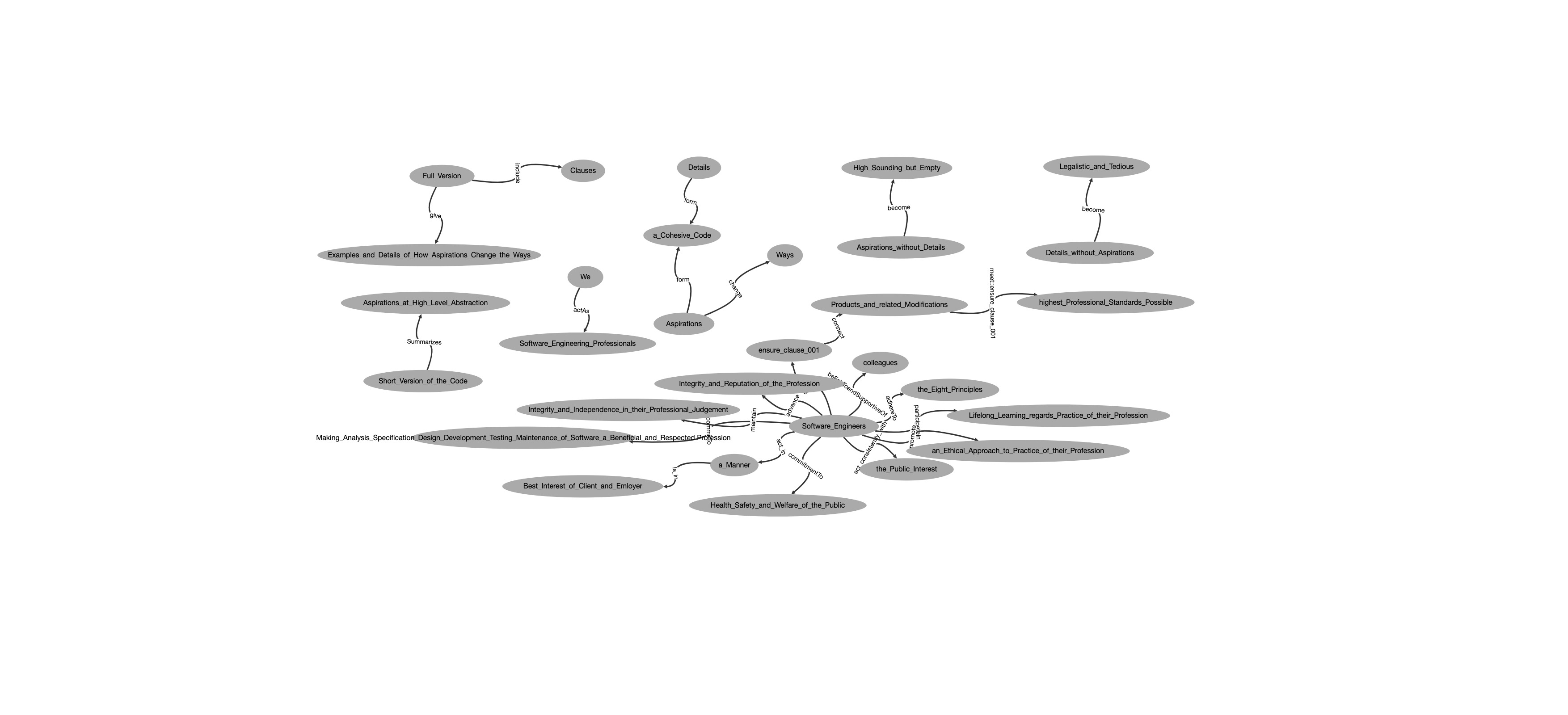}
		\caption{Ref--Long}\label{fig:og-a}
	\end{subfigure}\hfill
	\begin{subfigure}[b]{0.32\textwidth}
		\centering
		\includegraphics[width=\linewidth]{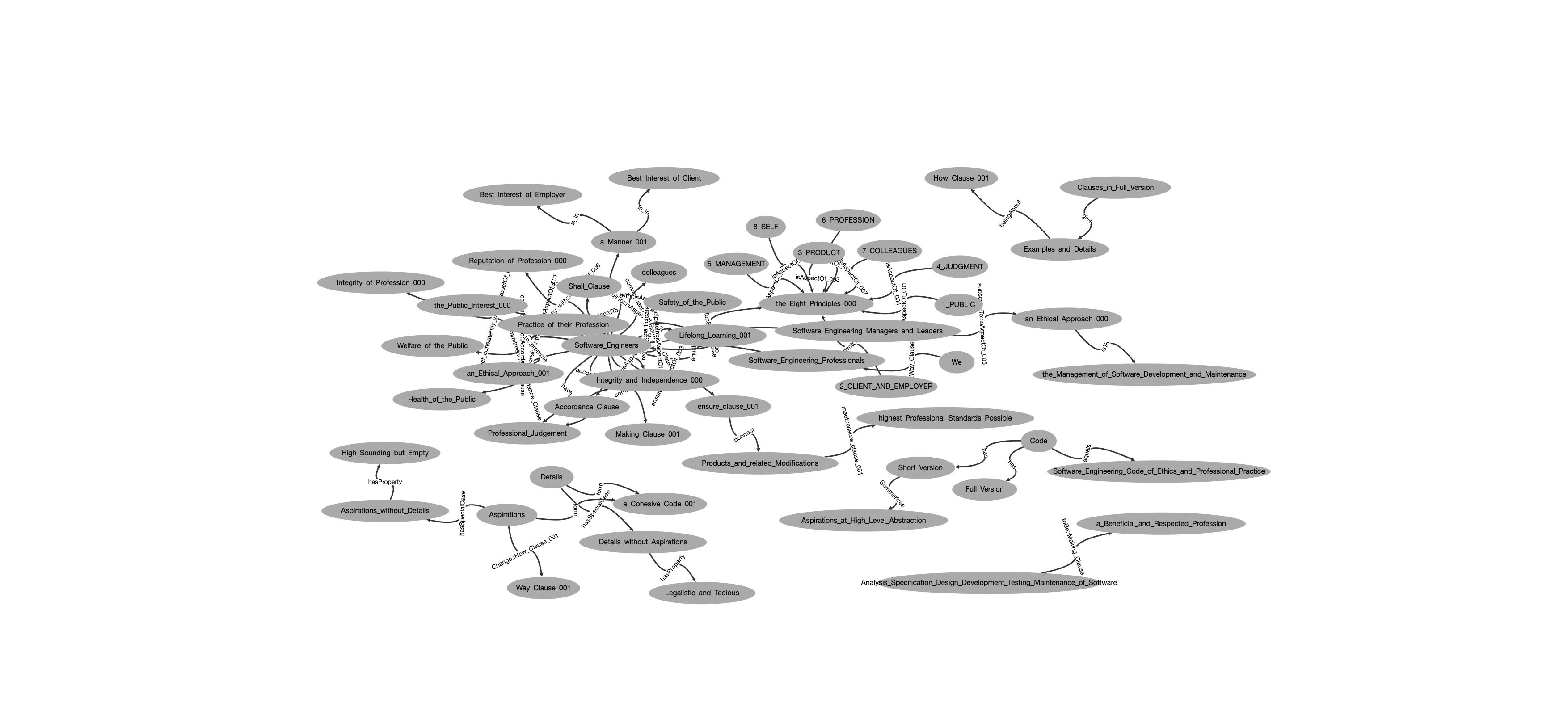}
		\caption{Ref--Medium}\label{fig:og-b}
	\end{subfigure}\hfill
	\begin{subfigure}[b]{0.32\textwidth}
		\centering
		\includegraphics[width=\linewidth]{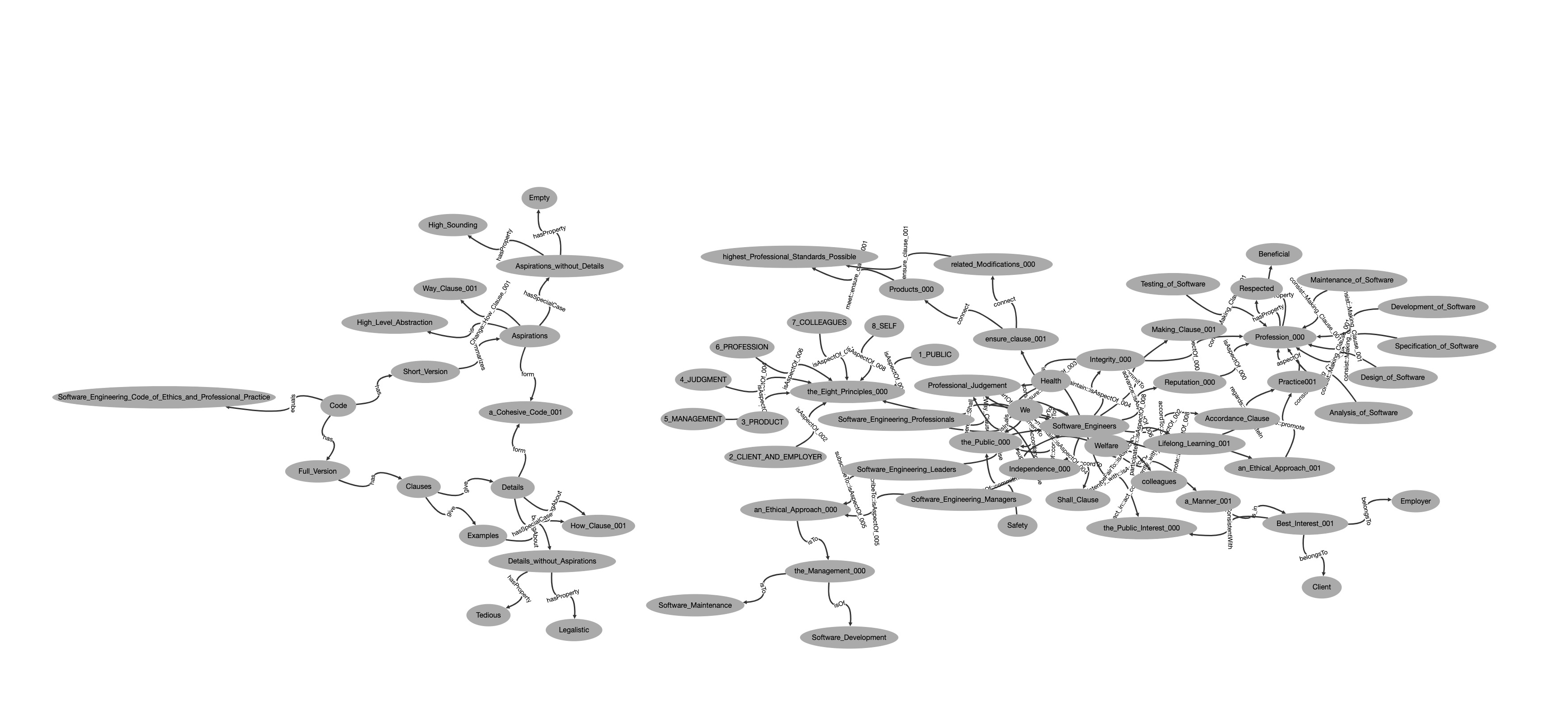}
		\caption{Ref--Short}\label{fig:og-c}
	\end{subfigure}
	
	\vspace{6pt}
	
	\makebox[\textwidth][c]{%
		\begin{subfigure}[b]{0.32\textwidth}\centering
			\includegraphics[width=\linewidth]{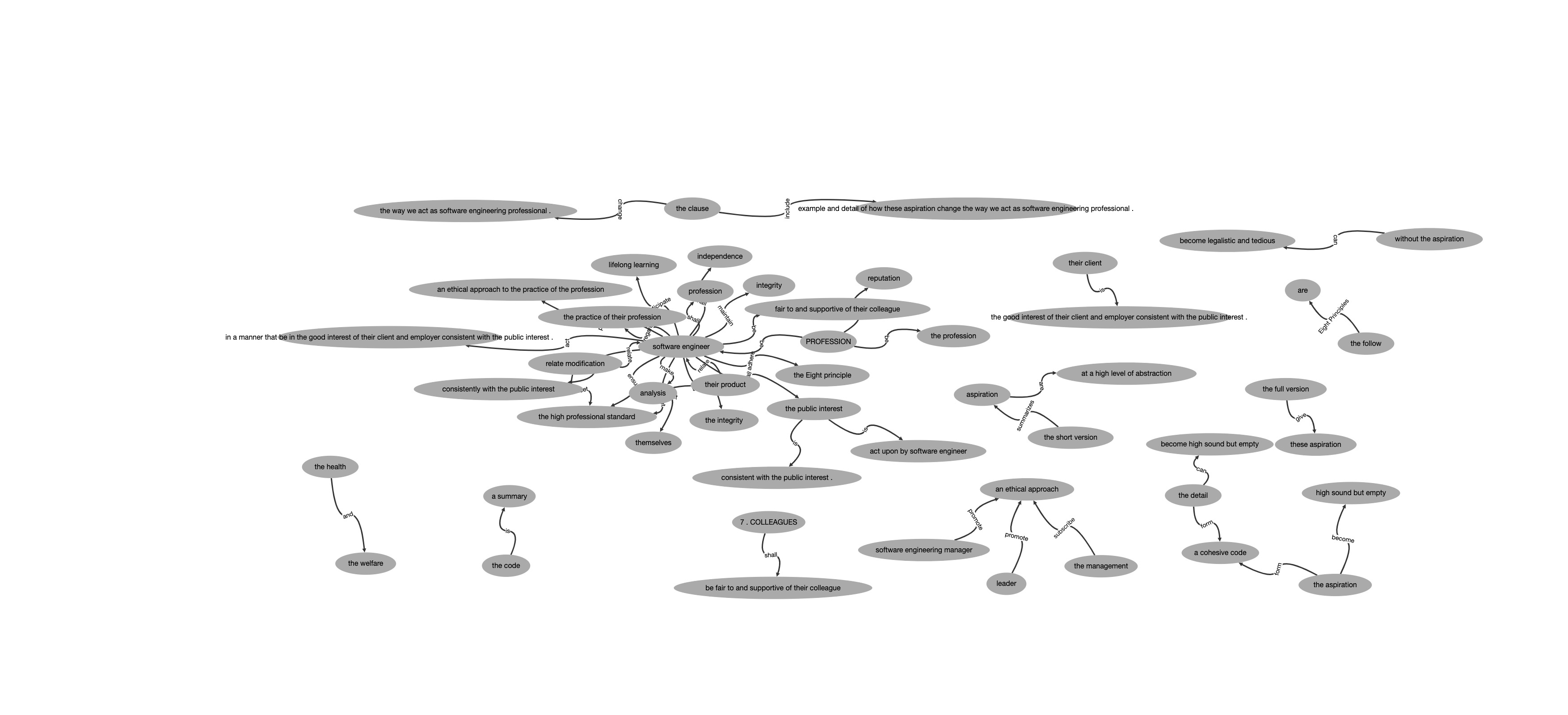}
			\caption{Pred}\label{fig:og-d}
		\end{subfigure}\hspace{0.04\textwidth}
		\begin{subfigure}[b]{0.32\textwidth}\centering
			\includegraphics[width=\linewidth]{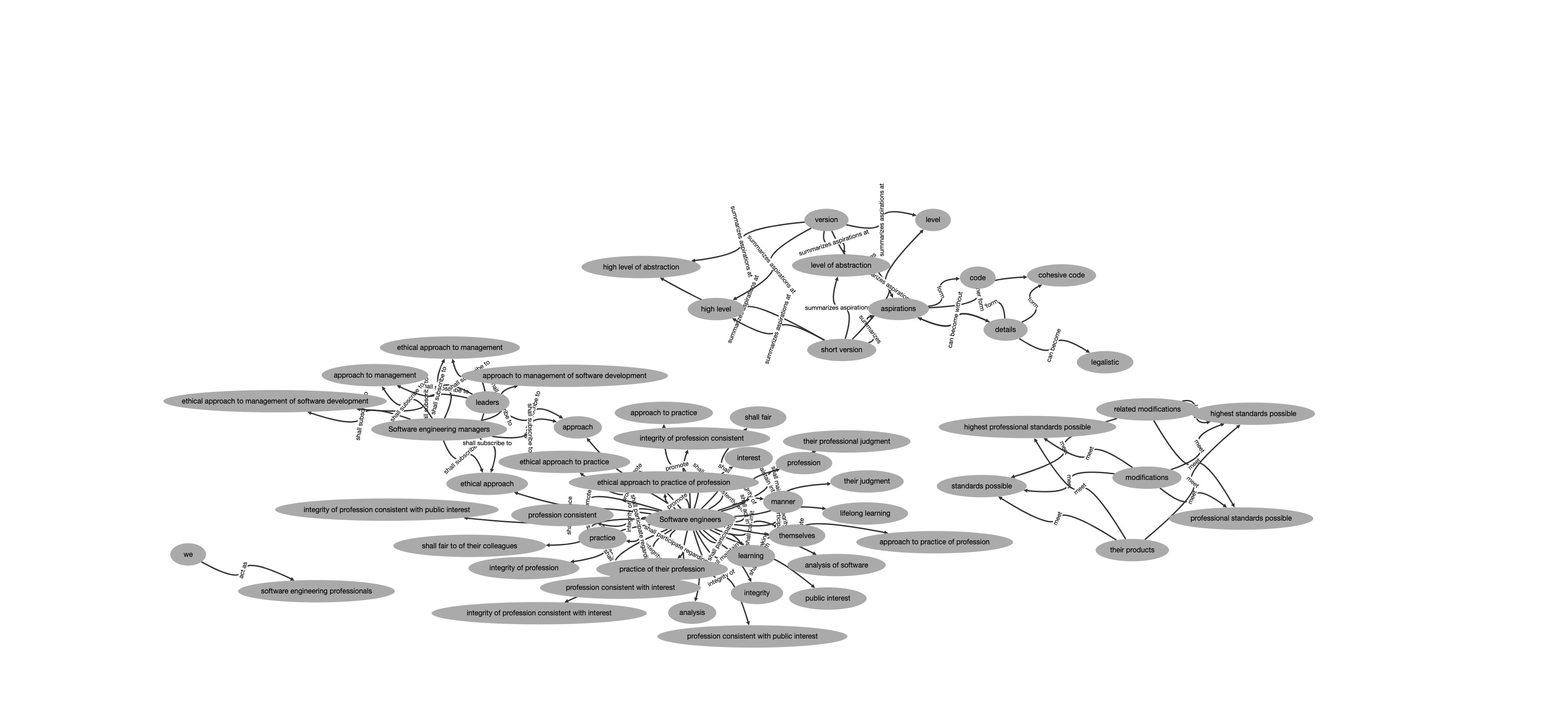}
			\caption{Pred--OpenIE}\label{fig:og-e}
		\end{subfigure}
	}
	
	\caption{Visualized Ontology Scaffolds.}
	\Description{Five ontology graph panels: Ref-Long, Ref-Medium, Ref-Short on the first row; Pred and Pred-OpenIE centered on the second row.}
	\label{fig:ontology-panels}
\end{figure*}

Three expert-annotated reference sets—Ref-Long, Ref-Medium, and Ref-Short—are constructed manually for the short version of the SECEPP. They enable us to evaluate the robustness of the method across various term granularities and levels of strictness, and their visualization corresponds to Figure \ref{fig:ontology-panels} panels (a), (b), and (c). Ref-Short preserves fine-grained clauses (higher density, recall-oriented as recall is more difficult to achieve and thus more thoroughly tested), Ref-Medium merges minor edges of Ref-Short, and Ref-Long preserves longer clauses and core relations (precision-oriented).

Table \ref{tab:dataset-stats} includes the numbers of Nodes, Triples, and Islands for the three reference sets. As expected, Ref-Short has the largest graph (68 nodes/83 triples), with fewer islands. Ref-Medium is more fragmented (26 islands) due to pruning connections, providing a balanced middle ground. Ref-Long is the most compact (32 nodes/25 triples).

\begin{table}[t]
	\caption{Counts for Pred (system outputs) and expert-annotated reference sets (Short/Medium/Long).}
	\Description{Counts of nodes, triples, and islands for Pred and three reference sets.}
	\label{tab:dataset-stats}
	\centering
	\small
	\begin{threeparttable}
		\begin{tabular}{@{}lrrr@{}}
			\toprule
			Dataset      & \#Nodes & \#Triples & \#Islands \\
			\midrule
			\textbf{Pred-LLM} (average)         & 55    & 52      & 12 \\
			Pred-openIE         & 53    & 74      & 4 \\
			Ref--Short  & 68     & 83       & 6 \\
			Ref--Medium & 72     & 54      & 26 \\
			Ref--Long   & 32    & 25      & 8 \\
			\bottomrule
		\end{tabular}
		\begin{tablenotes}[para,flushleft]
		\end{tablenotes}
	\end{threeparttable}
\end{table}

\begin{figure*}[t]
	\centering
	\includegraphics[width=0.6\textwidth]{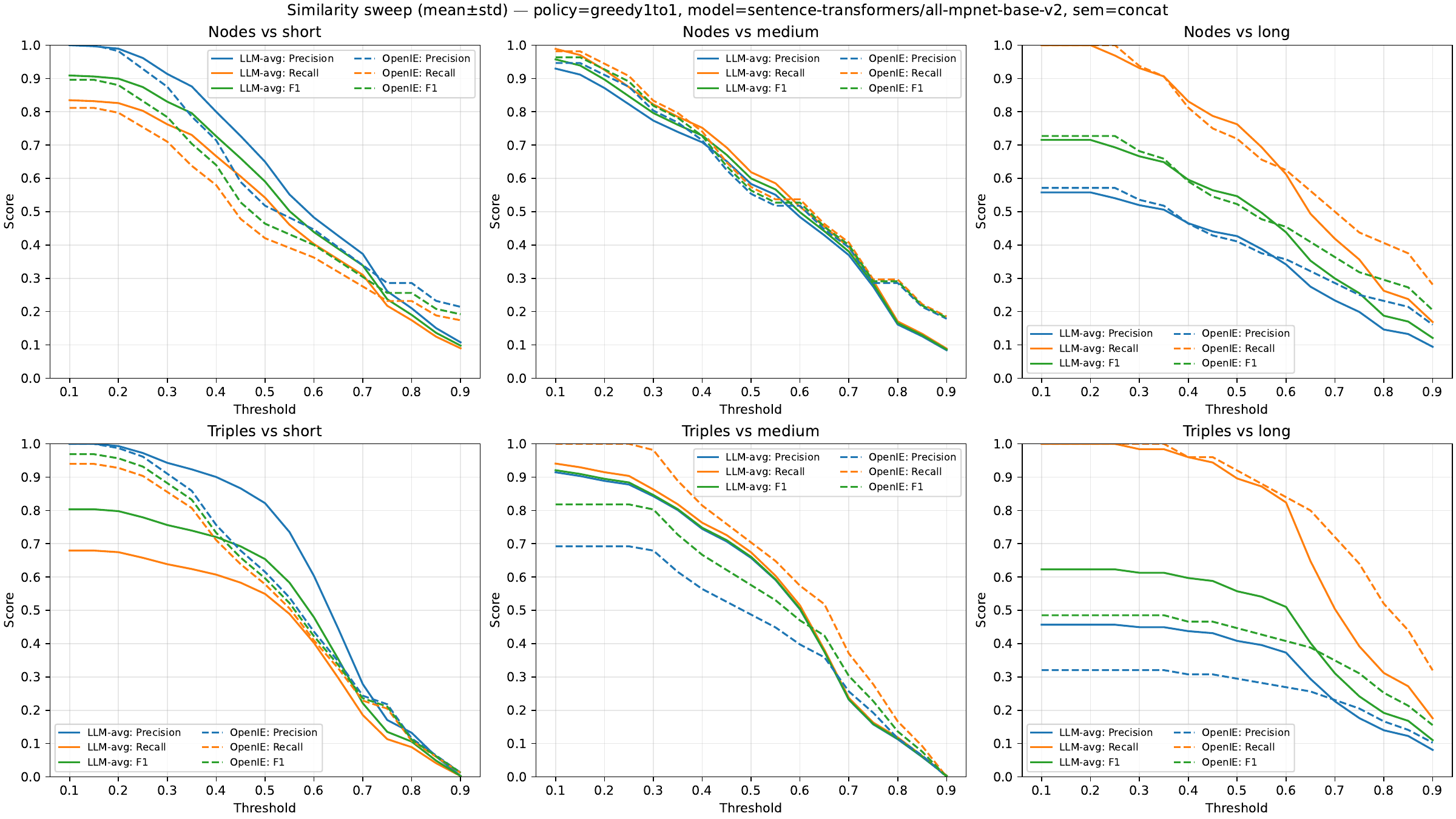}
	\caption{Similarity-threshold Sweep: Top row shows node metrics (Precision/Recall/F1) vs. short/medium/long reference; bottom row shows triple metrics.}
	\Description{Six subplots showing precision, recall, and F1 across thresholds for nodes and triples, comparing short, medium, and long reference sets.}
	\label{fig:sweep-curves}
\end{figure*}

\section{Experiments and Preliminary Results}
\label{sec:experiments}

\subsection{Systems Compared}
Pred-LLM is the result from this research–Figure \ref{fig:ontology-panels} panel (d) is one representative output out of five runs, the mean performance of which is reported in Figure \ref{fig:sweep-curves}. An open-source 7B LLM (Mistral-7B-Instruct-v0.1 \cite{jiang2023mistral7b}) is used in this study of open-set relation generation. Code and data are available at \url{https://github.com/songhui01/aog-ses.git}.

Pred-openIE is a sentence-level OpenIE baseline–an example is shown in Figure \ref{fig:ontology-panels} panel (e). Stanford OpenIE \cite{angeli2015leveraging} implemented in Stanford CoreNLP \cite{manning2014stanford} is employed.

\subsection{Evaluation Protocol}
Fully automated outputs (no post-editing) are evaluated against three \textit{ref-standard} reference sets for the short SECEPP:
\textit{Ref-Short}, \textit{Ref-Medium}, \textit{Ref-Long} (Figure ~\ref{fig:ontology-panels}; stats in Table~\ref{tab:dataset-stats}).

Quality is assessed at the \textit{node} and \textit{triple} levels via an embedding-similarity threshold sweep $\tau \in \{0.10, 0.15, \dots, 0.90\}$;  Precision, Recall and F1 are reported as curves (Figure \ref{fig:sweep-curves}, top: nodes; bottom: triples; columns: Short/Medium/Long). Matches require the similarity of the concatenation of subject, relation, and object (a single space delimiteris used) to pass $\tau$ (greedy one-to-one alignment).

\subsection{Key Findings}
(1) Recall is highest on \textit{Ref-Long} and lowest on \textit{Ref-Short}, confirming the \textit{recall-stress} hypothesis for dense reference sets. The \textit{Ref-Long} is a compact backbone, and it eases recall (fewer edges), but any extra predicted edges not in the backbone are counted as false positives, lowering precision.\\
(2) \textit{Pred-LLM}  often yields higher precision than OpenIE across $\tau$, and only when $\tau\!>\!0.5$, the recall gap narrows, which makes sense as \textit{Pred-LLM}'s result has a much smaller number than \textit{Pred-openIE}'s result ($42$ vs.\ $74$ under Ref-Short) and spurious/duplicate triples from OpenIE are filtered.\\
(3) An additional observation from the experimental checks is that when $\tau$ reaches around 0.6, the triples generally stabilize and become reliable. The supporting evidence could not be included in this short paper, but this preliminary finding motivates further investigation of the optimal $\tau$ value as the key indicator for this type of study.\\
(4 )A 7B open-source LLM is already \textit{competitive} with OpenIE—\textit{higher precision} on both nodes and triples, and \textit{comparable recall} on stricter reference sets (when $\tau\!>\!0.5$)—while recall on \textit{Ref-Short} remains the main headroom to be improved. \\
(5) Triple matching is strictly harder than node matching: errors in any of head/relation/tail cause a miss, so generally both precision and recall drop at the triple level.

\subsection{Discussion}
The results suggest that a 7B open-source LLM can already be competitive with OpenIE in precision for node and triple extraction. This indicates that stronger LLMs—either with larger parameter sizes or fine-tuned on domain-specific data—may further improve extraction quality.

However, this preliminary study has several limitations. First, it evaluates the workflow using a single LLM and thus lacks comparative experiments across LLM families and scales, as well as across backbone variants such as dense vs. MoE. Second, the reference datasets are manually constructed by a single annotator (domain expert) at three different granularities to support systematic evaluation in the absence of existing gold sets for this domain/task; consequently, inter-annotator agreement (IAA) is not reported. These labels are intended for formative evaluation rather than a community benchmark. In future work, we will include multiple annotators and report IAA to improve label reliability and support the construction of a public benchmark dataset.

\section{Conclusion / Future Work}
\label{sec:conclusion}

This work initiates LLM-assisted ontology generation for software engineering standards, proposes a fully automated workflow, constructs three expert-annotated reference sets (single annotator in this priliminary study) for the short version, and evaluates with threshold-sweep metrics at node and triple levels. Preliminary results show that a 7B open-source LLM is competitive with a Stanford OpenIE baseline (often higher precision, comparable recall on stricter reference sets). 

Next, we will work on the improvement of the results, extend from the short to the full (long) version and to additional SES documents; improve recall via cross-sentence (coreference) handling, and small-scope fine-tuning; and perform the later ontology-transformation phases to produce and release a formal OWL~2 ontology and public benchmarks.

\section{Acknowledgements}
This material is based upon work supported by the National Science Foundation under Grant No. 2451397.

AI assistants (OpenAI’s GPT-4.x and Google’s Gemini suggestions embedded in Google Colab) were used for rapid prototyping during the experiments. No AI was used to fabricate data or results. The code was comprehensively reviewed and revised by the author. AI tools were also used to rephrase some passages for conciseness and grammatical correctness.
\bibliography{mybibtex}

@article{bedini2007automatic,
	title={Automatic ontology generation: State of the art},
	author={Bedini, Ivan and Nguyen, Benjamin},
	journal={PRiSM Laboratory Technical Report. University of Versailles},
	pages={1--15},
	year={2007}
}

@article{karanikolas2025strengths,
	title={Strengths and Weaknesses of LLM-Based and Rule-Based NLP Technologies and Their Potential Synergies},
	author={Karanikolas, Nikitas N and Manga, Eirini and Samaridi, Nikoletta and Stergiopoulos, Vaios and Tousidou, Eleni and Vassilakopoulos, Michael},
	journal={Electronics},
	volume={14},
	number={15},
	pages={3064},
	year={2025},
	publisher={MDPI}
}

@article{zhang2024mm,
	title={Mm-llms: Recent advances in multimodal large language models},
	author={Zhang, Duzhen and Yu, Yahan and Dong, Jiahua and Li, Chenxing and Su, Dan and Chu, Chenhui and Yu, Dong},
	journal={arXiv preprint arXiv:2401.13601},
	year={2024}
}

@article{kommineni2024human,
	title={From human experts to machines: An LLM supported approach to ontology and knowledge graph construction},
	author={Kommineni, Vamsi Krishna and K{\"o}nig-Ries, Birgitta and Samuel, Sheeba},
	journal={arXiv preprint arXiv:2403.08345},
	year={2024}
}

@inproceedings{opasjumruskit2022ontohuman,
	title={OntoHuman: ontology-based information extraction tools with human-in-the-loop interaction},
	author={Opasjumruskit, Kobkaew and B{\"o}ning, Sarah and Schindler, Sirko and Peters, Diana},
	booktitle={International Conference on Cooperative Design, Visualization and Engineering},
	pages={68--74},
	year={2022},
	organization={Springer}
}

@article{tsaneva2024enhancing,
	title={Enhancing human-in-the-loop ontology curation results through task design},
	author={Tsaneva, Stefani and Sabou, Marta},
	journal={ACM Journal of Data and Information Quality},
	volume={16},
	number={1},
	pages={1--25},
	year={2024},
	publisher={ACM New York, NY}
}

@article{lo2024end,
	title={End-to-end ontology learning with large language models},
	author={Lo, Andy and Jiang, Albert Q and Li, Wenda and Jamnik, Mateja},
	journal={Advances in Neural Information Processing Systems},
	volume={37},
	pages={87184--87225},
	year={2024}
}

@article{kibriya2024privacy,
	title={Privacy issues in large language models: a survey},
	author={Kibriya, Hareem and Khan, Wazir Zada and Siddiqa, Ayesha and Khan, Muhammad Khurram},
	journal={Computers and Electrical Engineering},
	volume={120},
	pages={109698},
	year={2024},
	publisher={Elsevier}
}

@misc{acm1999code,
	title        = {The Software Engineering Code of Ethics and Professional Practice},
	author       = {{ACM/IEEE-CS joint task force on Software Engineering Ethics and Professional Practices (SEEPP)}},
	year         = {1999},
	note         = {Copyright (c) 1999 by the Association for Computing Machinery, Inc. and the Institute for Electrical and Electronics Engineers, Inc.},
	howpublished = {https://www.acm.org/code-of-ethics/software-engineering-code},
}

@article{gong2021prompt,
	title={Prompt-based zero-shot relation extraction with semantic knowledge augmentation},
	author={Gong, Jiaying and Eldardiry, Hoda},
	journal={arXiv preprint arXiv:2112.04539},
	year={2021}
}

@article{han2018fewrel,
	title={FewRel: A large-scale supervised few-shot relation classification dataset with state-of-the-art evaluation},
	author={Han, Xu and Zhu, Hao and Yu, Pengfei and Wang, Ziyun and Yao, Yuan and Liu, Zhiyuan and Sun, Maosong},
	journal={arXiv preprint arXiv:1810.10147},
	year={2018}
}

@article{honnibal2020spacy,
	title={spaCy: Industrial-strength natural language processing in python},
	author={Honnibal, Matthew and Montani, Ines and Van Landeghem, Sofie and Boyd, Adriane and others},
	year={2020},
	publisher={Zenodo, Honolulu, HI, USA}
}

@inproceedings{angeli2015leveraging,
	title={Leveraging linguistic structure for open domain information extraction},
	author={Angeli, Gabor and Premkumar, Melvin Jose Johnson and Manning, Christopher D},
	booktitle={Proceedings of the 53rd Annual Meeting of the Association for Computational Linguistics and the 7th International Joint Conference on Natural Language Processing (Volume 1: Long Papers)},
	pages={344--354},
	year={2015}
}

@inproceedings{manning2014stanford,
	title={The Stanford CoreNLP natural language processing toolkit},
	author={Manning, Christopher D and Surdeanu, Mihai and Bauer, John and Finkel, Jenny Rose and Bethard, Steven and McClosky, David},
	booktitle={Proceedings of 52nd annual meeting of the association for computational linguistics: system demonstrations},
	pages={55--60},
	year={2014}
}

@misc{jiang2023mistral7b,
	title={Mistral 7B}, 
	author={Albert Q. Jiang and Alexandre Sablayrolles and Arthur Mensch and Chris Bamford and Devendra Singh Chaplot and Diego de las Casas and Florian Bressand and Gianna Lengyel and Guillaume Lample and Lucile Saulnier and Lélio Renard Lavaud and Marie-Anne Lachaux and Pierre Stock and Teven Le Scao and Thibaut Lavril and Thomas Wang and Timothée Lacroix and William El Sayed},
	year={2023},
	eprint={2310.06825},
	archivePrefix={arXiv},
	primaryClass={cs.CL},
	url={https://arxiv.org/abs/2310.06825}, 
}

@article{yue2023csm,
	title={CSM-HR: A Context Modeling Framework in Supporting Reasoning Automation for Interoperable Intelligent Systems and Privacy Protection},
	author={Yue, Songhui and Hong, Xiaoyan and Smith, Randy K},
	journal={arXiv e-prints},
	pages={arXiv--2308},
	year={2023}
}

@inproceedings{yue2021applying,
	title={Applying context state machines to smart elevators: Design, implementation and evaluation},
	author={Yue, Songhui and Smith, Randy K},
	booktitle={2021 IEEE Symposium Series on Computational Intelligence (SSCI)},
	pages={1--9},
	year={2021},
	organization={IEEE}
}

@article{al2020automatic,
	title={Automatic ontology construction from text: a review from shallow to deep learning trend},
	author={Al-Aswadi, Fatima N and Chan, Huah Yong and Gan, Keng Hoon},
	journal={Artificial Intelligence Review},
	volume={53},
	number={6},
	pages={3901--3928},
	year={2020},
	publisher={Springer}
}

@article{navarro2020automated,
	title={Automated ontology extraction from unstructured texts using deep learning},
	author={Navarro-Almanza, Ra{\'u}l and Ju{\'a}rez-Ram{\'\i}rez, Reyes and Licea, Guillermo and Castro, Juan R},
	journal={Intuitionistic and Type-2 fuzzy logic enhancements in neural and optimization algorithms: Theory and applications},
	pages={727--755},
	year={2020},
	publisher={Springer}
}

@article{elnagar2022automatic,
	title={An automatic ontology generation framework with an organizational perspective},
	author={Elnagar, Samaa and Yoon, Victoria and Thomas, Manoj A},
	journal={arXiv preprint arXiv:2201.05910},
	year={2022}
}

@article{leshcheva2022method,
	title={A method of semi-automated ontology population from multiple semi-structured data sources},
	author={Leshcheva, Irina and Begler, Alena},
	journal={Journal of Information Science},
	volume={48},
	number={2},
	pages={223--236},
	year={2022},
	publisher={Sage Publications Sage UK: London, England}
}

@inproceedings{babaei2023llms4ol,
	title={LLMs4OL: Large language models for ontology learning},
	author={Babaei Giglou, Hamed and D’Souza, Jennifer and Auer, S{\"o}ren},
	booktitle={International Semantic Web Conference},
	pages={408--427},
	year={2023},
	organization={Springer}
}

@inproceedings{lippolis2025ontology,
	title={Ontology generation using large language models},
	author={Lippolis, Anna Sofia and Saeedizade, Mohammad Javad and Keskis{\"a}rkk{\"a}, Robin and Zuppiroli, Sara and Ceriani, Miguel and Gangemi, Aldo and Blomqvist, Eva and Nuzzolese, Andrea Giovanni},
	booktitle={European Semantic Web Conference},
	pages={321--341},
	year={2025},
	organization={Springer}
}

@article{neuhaus2023ontologies,
	title={Ontologies in the era of large language models--a perspective},
	author={Neuhaus, Fabian},
	journal={Applied ontology},
	volume={18},
	number={4},
	pages={399--407},
	year={2023},
	publisher={SAGE Publications Sage UK: London, England}
}

@inproceedings{omelayenko2001learning,
	title={Learning of Ontologies from the Web: the Analysis of Existent Approaches.},
	author={Omelayenko, Borys},
	booktitle={WebDyn@ ICDT},
	pages={16--25},
	year={2001},
	organization={Citeseer}
}

@article{lakzaei2021ontology,
	title={Ontology learning from relational databases},
	author={Lakzaei, Batool and Shamsfard, Mehrnoush},
	journal={Information Sciences},
	volume={577},
	pages={280--297},
	year={2021},
	publisher={Elsevier}
}

@inproceedings{maedche2000semi,
	title={Semi-automatic engineering of ontologies from text},
	author={Maedche, Alexander and Staab, Steffen},
	booktitle={Proceedings of the 12th international conference on software engineering and knowledge engineering},
	pages={231--239},
	year={2000},
	organization={Chicago, IL, USA}
}
\bibliographystyle{ieeetr}

\end{document}